\documentclass[aps,showpacs,prd,twocolumn]{revtex4-1}
\usepackage{amsfonts}
\usepackage{amssymb}
\usepackage{amsmath}
\usepackage{color}
\usepackage[usenames,dvipsnames]{xcolor}
\usepackage{float}
\usepackage{accents}
\usepackage{graphicx}
\usepackage{epstopdf}
\usepackage[colorlinks=true,pdfstartview=FitV,linkcolor=blue,citecolor=blue,urlcolor=blue,breaklinks=true]{hyperref}
\usepackage{ulem}

\begin{document}

\title{Self-dual solitons in a Maxwell-Chern-Simons baby Skyrme model}
\author{Rodolfo Casana}
\email{rodolfo.casana@gmail.com}
\author{Andr\'{e} C. Santos}
\email{andre$\_$cavs@hotmail.com}
\author{Claudio F. Farias}
\email{cffarias@gmail.com}
\author{Alexsandro L. Mota}
\email{lucenalexster@gmail.com}
\affiliation{Departamento de F\'{\i}sica, Universidade Federal do Maranh\~{a}o,
65080-805, S\~{a}o Lu\'{\i}s, Maranh\~{a}o, Brazil.}

\begin{abstract}
We have studied the existence de self-dual solitons in a gauged version of
the baby Skyrme model in which the gauge field dynamics is governed by the Maxwell-Chern-Simons action. For such a purpose, we have developed a detailed implementation of the Bogomol'nyi-Prasad-Sommerfield formalism providing the self-dual equations whose solutions saturate the energy lower bound. Such a bound related to the topological charge of the Skyrme field becomes quantized whereas both the total magnetic flux and the total electrical charge are not. We have found two types of self-dual {\color{black}Skyrme field profiles: the first is described by a solution which} decays following an exponential-law ($e^{-\alpha r^2}$\!, $\alpha>0$); the second {\color{black}is portrayed by a solution having} a power-law decay ($r^{-\beta}$\!, $\beta>0$). {\color{black}{\color{black}On other hand, in both cases the asymptotic behavior of the gauge field  is similar to the one presented {\color{black} in the context of the} {\color{black}Abelian Higgs models} describing Abrikosov-Nielsen-Olesen charged vortices.}  Other interesting feature we highlight is the localized magnetic {\color{black}flux inversion,} a property does not observed in others gauged baby Skyrme models {\color{black}already studied in literature.} Numerical results are presented  for rotationally symmetrical field configurations by remarking some of its essential features.}
\end{abstract}

\maketitle

\section{Introduction}

The Skyrme model \cite{skyrme} is a non-linear field theory, which is
originally defined in  {\color{black}$(3+1)$} dimensions and whose soliton solutions are called Skyrmions, {\color{black}it has been} a prolific subject in several branches of physics. Initially, it was proposed as an effective field theory for nuclear phenomena that avoid some technical difficulties present in its underlining theory, more specifically the Quantum Chromodynamics. This model was reasonably successful in obtaining several hadrons and nucleons properties \cite{adkins,adkins2, braaten,manko}, with the latter emerging as topological soliton solutions called Skyrmions. {\color{black}Further, recently it has} found an exciting research field in the realm of condensed matter physics. The Skyrme model has been studied for physical systems such as liquid Helium \cite{volovik,volovik2}, quantum hall effect \cite{Soundhi,sarma,Walet}, Bose-Einstein condensates \cite{usama} and chiral nematic liquid crystals \cite{jun}. However, most promising results were
found in magnetic materials \cite{muhlbauer,yi,yu} and superconductors \cite{garaud,garaud2,zyuzin,winyard,vadimov,bogdanov3,bogdanov4,bogdanov5,robler}. The first one provides some
theoretical and experimental realizations that can yield important
technological applications such as data storage and spintronic.

The $(2+1)$-dimensional version of the full model \cite{skyrme} is called
baby Skyrme model \cite{piette} being described by the Lagrangian
density
\begin{equation}
\mathcal{L}=\frac{\nu^{2}}{2}\partial _{\mu }\vec{\phi}\cdot \partial
^{\mu }\vec{\phi}-\frac{\lambda ^{2}}{4}(\partial _{\mu }\vec{\phi}\times
\partial _{\nu }\vec{\phi})^{2}-V(\phi _{n}).  \label{lagbsk0}
\end{equation}%
The first contribution is the well-known nonlinear $\sigma $-model term. The
second term {\color{black}is the counterpart of} the Skyrme term in Ref. \cite{skyrme}. The last term, $V(\phi _{n})\equiv V(\vec{n}\cdot \vec{\phi})$, is an appropriate self-interacting potential added with aim to guarantee the
stability of the soliton solutions \cite{Leese}. The vector $\vec{\phi}$ is
the Skyrme field denoting a triplet of real scalar fields $\vec{\phi}%
=\left(\phi _{1},\phi _{2},\phi _{3}\right) $ satisfying the constraint $%
\vec{\phi}\cdot \vec{\phi} =\phi _{1}^{2}+\phi _{2}^{2}+\phi _{3}^{2}=1$,
which stands an unit two-sphere (denoted by $\mathbb{S}^{2}$), and the
unitary vector $\vec{n}\in \mathbb{S}^{2}$ provides a preferred direction in
the internal space. The $\sigma $-model and Skyrme terms are invariant under
the $SO(3)$ global symmetry whereas the potential breaks partially it but
preserving the $U(1)$ subgroup. The potential has an unique vacuum
configuration and must satisfy the condition $V(\phi _{n})\rightarrow 0$
when $\phi _{n}\rightarrow 1$.

In absence of the $\sigma$-model term, the resulting is the named \textit{%
restricted} baby Skyrme model which was firstly considered in Ref. \cite%
{gisiger}. The Bogomol'nyi structures [obtained via the
Bogomol'nyi-Prasad-Sommerfield (BPS) formalism] in this model were
investigated in Ref. \cite{adam}, there has been found the energy lower bond
(Bogomol'nyi limit or BPS bound) and the respective self-dual or BPS
equations satisfied by the soliton configurations saturating such a bound.

A natural physical extension in the study of the baby Skyrme model is the
possibility of its coupling to the electromagnetic field in order to
investigate their electric and/or magnetic properties. It is achieved by
promoting the $U(1)$ global symmetry to be a local gauge symmetry by means
of the introduction of {\color{black} an} Abelian gauge field, whose dynamics can be governed
solely by the Maxwell term \cite{gladikowski,adam4, samoilenka2, samoilenka4}
or the Chern-Simons term \cite{Casana01}, or also by both the Maxwell and
the Chern-Simons terms \cite{Loginov,samoilenka, Navarro,Navarro2}.

Even though of investigations involving BPS structures in the gauged
versions of the nonlinear $\sigma$-model (for example, when it is minimally
coupled to the Maxwell term \cite{schroers} or the Chern-Simons term \cite%
{Ghosh}, or still both the Maxwell and the Chern-Simons terms \cite{CLee})
has been successful, similar results by applying the BPS formalism have
never been obtained for a baby Skyrme model. Such a limitation can be
partially circumvented by adopting gauged versions of the restricted baby
Skyrme model. Thus, BPS solitons has been obtained in a gauged restricted
model whose gauge field dynamics is governed by the Maxwell term \cite{adam2}%
. All these results have inspired the development a wide range of applications,
including topological phase transitions \cite{adam4}, Bogomol'nyi equations
based in the strong necessary conditions limit \cite{stepien}, gauged BPS
baby skyrmions with quantized magnetic flux \cite{adam5} and even in
supersymmetry \cite{susy1, susy2, susy3,susy4,Queiruga} and gravitational
theories \cite{gravity}.

{\color{black}Recently {\color{black} was well established} the existence of self-dual configurations in a generalized Chern-Simons baby Skyrme model {\color{black}by means of a} successful implementation of the BPS technique  \cite{Casana01}.} {\color{black}Further, in Refs. \cite{Loginov, samoilenka} has been studied some properties of the solitonic solutions emerging in the Maxwell-Chern-Simons baby Skyrme model, nevertheless, still  remains open the obtaining of self-dual solitons via direct application of the BPS formalism.
On the other side, the existence of BPS structures  in gauged baby Skyrme models including the Chern-Simons term also has been revealed by using supersymmetry techniques \cite{Queiruga}, specifically,  these models possess  $N=2$ SUSY extension.  The matching of the BPS structures arising via the direct use of the BPS technique and SUSY formalism in the Chern-Simons baby Skyrme model has been discussed in \cite{Casana01}.}

{\color{black}Our aim is the construction of a BPS description of the solitonic solutions (and their main features) emerging in a gauged baby Skyrme model whose gauge field dynamic is governed by the Maxwell-Chern-Simons action.} The manuscript is divided as follows. {\color{black}In Secs. \ref{themodel} and \ref{themodel1},} we present the essential features of  the Maxwell-Chern-Simons restricted baby Skyrme model {\color{black}and the necessity to introduce an auxiliary dynamical field} with the aim to gain a BPS model. Immediately, the developing of the BPS formalism provides the BPS potential, the self-dual {\color{black}equations and} the Bogomol'nyi bound for the total energy. In Sec. \ref{bpssolutions}, we restrict our analysis to the study of rotationally symmetric solutions by discussing the boundary conditions and obtaining the physical quantities, namely, the magnetic flux and the electrical charge. In Sec. \ref{numerical},  we depict some relevant profiles and discuss their quantitative and qualitative features.
In Sec. \ref{conclusion}, we present our remarks and conclusions.

\section{Maxwell-Chern-Simons restricted baby Skyrme model\label{themodel}}

The gauged version of the baby Skyrme model which consider the Skyrme field
minimally coupled to the Maxwell-Chern-Simons field \cite{Loginov,samoilenka}
is described by the following Lagrangian density
\begin{eqnarray}
L &=& E_{0}\int d^{2}x\left[ -\frac{1}{4g^{2}}F_{\mu \nu }^{2}-\frac{\kappa
}{4g^{2}}\epsilon ^{\rho \mu \nu }A_{\rho }F_{\mu \nu }\right.  \notag \\%
[0.2cm]
&& \left. +\frac{\nu^{2}}{2}(D_{\mu }\vec{\phi})^{2} -\frac{\lambda ^{2}%
}{4} (D_{\mu }\vec{\phi}\times D_{\nu } \vec{\phi})^{2} -V(\phi _{n})\right]
\text{,}  \label{L0}
\end{eqnarray}%
where {$D_{\mu }\vec{\phi}$ is the covariant derivative} \cite%
{gladikowski,schroers}
\begin{equation}
D_{\mu }\vec{\phi}=\partial _{\mu }\vec{\phi}+A_{\mu }\vec{n}\times \vec{\phi%
}\text{,}
\end{equation}%
defining the coupling between the Abelian gauge field $A_{\mu}$ and the
Skyrme field $\vec{\phi}$. The first contribution in Eq. (\ref{L0}) is the
Maxwell term with $F_{\mu\nu }= \partial _{\mu }A_{\nu}-\partial_{\nu}A_{\mu
}$ and $A_{\mu }$ being the $U(1)$ gauge field. The other contributions
listed in order are: the Chern-Simons term, the nonlinear $\sigma$-model
term, the Skyrme term, and the self-interacting potential. Besides, we have
extracted a common energy factor $E_{0}$ \cite{gladikowski,adam2} setting
the model energy scale which for our objectives we shall always consider $%
E_{0}=1$. Also, it will be assumed all the coupling constants are
non-negative quantities. Moreover, the Chern-Simons coupling constant $%
\kappa $ and the electromagnetic coupling $g$ have mass dimension $1$, the
Skyrme coupling constant $\lambda$ has mass dimension $-1$, and $\nu$
is dimensionless.

We restrict our study of the model (\ref{L0}) to the case without the $%
\sigma $-model term ($\nu=0$), thus, the starting point of our analysis
will be the following Lagrangian density
\begin{align}
\mathcal{L}&=-\frac{1}{4g^{2}}F_{\mu \nu }^{2}-\frac{\kappa }{4g^{2}}%
\epsilon ^{\rho \mu \nu }A_{\rho }F_{\mu \nu }  \notag \\[0.2cm]
&\quad-\frac{\lambda ^{2}}{4}(D_{\mu }\vec{\phi}\times D_{\nu }\vec{\phi}%
)^{2}-V\text{,}  \label{L02}
\end{align}
it can be named the Maxwell-Chern-Simons restricted Skyrme model.  The respective equation of motion of the gauge field reads
\begin{equation}
\partial _{\sigma }F^{\sigma \mu }-\frac{\kappa }{2}\epsilon ^{\mu \alpha
\beta }F_{\alpha \beta }=g^{2}\vec{n}\cdot \vec{J}^{\mu }\text{,}
\label{gauge2}
\end{equation}
and the one for the Skyrme field results
\begin{equation}
D_{\mu }\vec{J}^{\mu }=-\frac{\partial V}{\partial \phi _{n}}\vec{n}\times
\vec{\phi}.  \label{gauge1}
\end{equation}%
The conserved current density $j^{\mu} =\vec{n}\cdot \vec{J}^{\mu }$ is
defined in terms of the vector $\vec{J}^{\mu }$ given by
\begin{equation}
\vec{J}^{\mu }=\lambda ^{2}[\vec{\phi}\cdot (D^{\mu }\vec{\phi}\times D^{\nu
} \vec{\phi})](D_{\nu }\vec{\phi})\text{.}  \label{curr}
\end{equation}

Our effort will be focused to the study of stationary soliton solutions, this way, below we write the field equations ruling this regimen. Thus, from Eq. (\ref{gauge2}), the Gauss law reads,
\begin{equation}
\partial _{i}E_{i}-\kappa B=g^{2}j_{0}\text{,}  \label{gL}
\end{equation}%
with $j_{0}=-\lambda ^{2}A_{0}(\vec{n}\cdot \partial _{i}\vec{\phi})^{2}$,
the electric charge density. We have defined the electric field components
by $E_{i}=F_{0i}=-\partial _{i}A_{0}$ while the magnetic is given by $%
B=F_{12} =\epsilon _{ij}\partial _{i}A_{j}$. The Gauss law point out the
mixing of the electric and magnetic sectors, an effect due to the presence
of the Chern-Simons term. Clearly, Eq. (\ref{gL}) implies the field
configurations besides to possess a nonzero magnetic flux $\Phi $ they also
carry a nonzero total electric charge $\mathcal{Q}_{\text{em}}$. The
relation between both physical quantities is obtained by integrating (under
appropriated boundary conditions) the Gauss law,
\begin{equation}
\mathcal{Q}_{\text{em}}=-\frac{\kappa }{g^{2}}\Phi ,  \label{QQq}
\end{equation}%
where
\begin{equation}
\mathcal{Q}_{\text{em}}=\int {d^{2}x\,j_{0}},\ \ \Phi =\int d^{2}x\,B .
\end{equation}

For the stationary Amp\`{e}re law we get
\begin{equation}
\partial _{i}B-\kappa E_{i}+\lambda ^{2}g^{2}(\vec{n}\cdot \partial _{i}\vec{%
\phi})Q=0\text{.}  \label{aL}
\end{equation}%
{\color{black}Above, we have defined the quantity}
\begin{equation}
Q\equiv \vec{\phi}\cdot (D_{1}\vec{\phi}\times D_{2}\vec{\phi})\text{,}
\label{Q1a}
\end{equation}
{\color{black}which also can be expressed by}
\begin{equation}
Q=\vec{\phi}\cdot (\partial _{1}\vec{\phi}\times \partial _{2}\vec{\phi}%
)+\epsilon _{ij}A_{i}(\vec{n}\cdot \partial _{j}\vec{\phi})\text{.}
\label{Q1}
\end{equation}
The first term $\vec{\phi}\cdot (\partial _{1}\vec{\phi}\times \partial _{2} \vec{\phi})$ is related to the topological charge or topological degree ({\color{black} also named winding number}) of the Skyrme field \cite{Rajaraman},
\begin{equation}
\deg [\vec{\phi}]=-\frac{1}{4\pi }\!\int d^{2}x\;\vec{\phi}\cdot (\partial
_{1}\vec{\phi}\times \partial _{2}\vec{\phi})=k\text{,}  \label{wn}
\end{equation}%
being $k\in\mathbb{Z}\setminus 0$.

Similarly, the stationary equation of motion of the Skyrme field is given by
\begin{eqnarray}
0&=&\lambda ^{2}\epsilon _{ij}D_{i}(QD_{j}\vec{\phi})+\lambda ^{2}\partial
_{j}[A_{0}^{2}(\vec{n}\cdot \partial _{j}\vec{\phi})](\vec{n}\times \vec{\phi%
})  \notag \\[0.2cm]
& & +\frac{\partial V}{\mathcal{\partial }\phi _{n}}(\vec{n}\times \vec{\phi}%
)\text{.}  \label{sL}
\end{eqnarray}

Until now all researches about solitons solutions obtained from the
Lagrangian density (\ref{L02}) have not been able to engender self-dual or
BPS configurations. Then, in order to obtain such a BPS structures we
propose to next a modified version of the model (\ref{L02}) such that the
gauge field equation remains unchanged.

\section{Self-dual Maxwell-Chern-Simons baby Skyrme model\label{themodel1}}

\label{IIA}

The corresponding self-dual model is constructed by introducing a neutral
scalar field which besides to interact only with the Skyrme field also
modify the potential. Such a BPS Maxwell-Chern-Simons baby
Skyrme model is described by the following Lagrangian density
\begin{eqnarray}
\mathcal{L} &=&-\frac{1}{4g^{2}}F_{\mu \nu }^{2}-\frac{\kappa }{4g^{2}}%
\epsilon ^{\rho \mu \nu }A_{\rho }F_{\mu \nu }-\frac{\lambda ^{2}}{4}(D_{\mu
}\vec{\phi}\times D_{\nu }\vec{\phi})^{2}  \notag \\[0.2cm]
&&+\frac{1}{2g^{2}}\partial _{\mu }\Psi \partial ^{\mu }\Psi +\frac{\lambda
^{2}}{2}(\vec{n}\cdot D_{\mu }\vec{\phi})^{2}\Psi ^{2}-U\text{,}  \label{L03}
\end{eqnarray}%
where $\Psi$ is the neutral scalar field, and $U$ is the potential which now
is a function of both the variable $\phi_{n}$ and the neutral scalar field,
i.e., $U=U(\phi _{n},\Psi)$. The procedure used to achieve the model (\ref%
{L03}) by means of the introduction of a neutral scalar field with the aim
to attain a successful implementation of the Bogomol'nyi technique is
already well known in literature. It was firstly used in the context of
Maxwell-Chern-Simons-Higgs models \cite{Lee}. Similar approaches in subsequent investigations have also successfully implemented as in \cite{Kimm,Bolognesi} and in some Lorentz-violating scenarios \cite{Casana04,Casana05,Casana06,CMGC}. Further, it is important to mention the Lagrangian density (\ref{L03}) has a matching to a $\mathcal{N}=2$ SUSY model, such a property can be verified in Ref. \cite{Queiruga}.

Furthermore, it is worthwhile to point out the penultimate term in (\ref{L03}%
) can be expressed as
\begin{equation}
(\vec{n}\cdot D_{\mu }\vec{\phi})^{2}=(D_{\mu }\vec{\phi})^2 -(\vec{n}\times
D_{\mu }\vec{\phi})^2,
\end{equation}
{\color{black}which allows us to} write the Lagrangian density at form
\begin{eqnarray}
\mathcal{L} &=&-\frac{1}{4g^{2}}F_{\mu \nu }^{2}-\frac{\kappa }{4g^{2}}%
\epsilon ^{\rho \mu \nu }A_{\rho }F_{\mu \nu }  \notag \\[0.2cm]
&&-\frac{\lambda ^{2}}{4}(D_{\mu}\vec{\phi}\times D_{\nu }\vec{\phi})^{2} +%
\frac{1}{2g^{2}}\partial _{\mu }\Psi \partial ^{\mu }\Psi  \notag \\[0.2cm]
&&+\frac{\lambda ^{2}}{2} (D_{\mu }\vec{\phi})^2\Psi ^{2} -\frac{\lambda ^{2}%
}{2}(\vec{n}\times D_{\mu }\vec{\phi})^2\Psi ^{2}-U\text{.}  \label{L003}
\end{eqnarray}%
In other words, the model given in (\ref{L003}) can be considered a type of
generalized Maxwell-Chern-Simons-Skyrme model in $(2+1) $ dimensions,
including a $\sigma$-model like term, $(D_{\mu }\vec{\phi})^2\Psi ^{2}$, in
which the function $\Psi$ would play the role of the generalizing function.

As already mentioned, the gauge field equation corresponding to the model (%
\ref{L03}) is the same one of the model (\ref{L02}), i.e., the introduction
of the neutral scalar field $\Psi$ does not modify the gauge field equation
which is given by Eq. (\ref{gauge2}). However, the Skyrme field equation of
motion (\ref{gauge1}) suffers modification and now it reads
\begin{equation}
D_{\mu }\vec{J}^{\mu }=\left[-\lambda ^{2}\partial _{\mu }[(\vec{n}\cdot
D^{\mu }\vec{\phi})\Psi ^{2}]-\frac{\partial U}{\partial \phi _{n}}\right](%
\vec{n}\times \vec{\phi})\text{.}
\end{equation}%
The respective stationary version reads
\begin{eqnarray}
\frac{\partial U}{\mathcal{\partial }\phi _{n}}(\vec{n}\times \vec{\phi})
&=&\lambda ^{2}\partial _{i}[(\vec{n}\cdot \partial _{i}\vec{\phi})\left(
\Psi ^{2}-A_{0}^{2}\right) ](\vec{n}\times \vec{\phi})  \notag \\[0.2cm]
&&-\lambda ^{2}\epsilon _{ij}D_{i}(QD_{j}\vec{\phi})\text{.}  \label{Fs}
\end{eqnarray}

For last, the equation of motion of the neutral scalar field is
\begin{equation}
\partial _{\mu }\partial ^{\mu }\Psi -\lambda ^{2}g^{2}(\vec{n}\cdot D_{\mu }%
\vec{\phi})^{2}\Psi +g^{2}\frac{\partial U}{\partial \Psi }=0\text{.}
\end{equation}

In the next section, we will show how the BPS formalism is implemented. Along the procedure, the self-dual potential $U(\phi_{n},\Psi)$ is determined allowing to obtain the energy lower bound and the self-dual
equations satisfied by the solitonic configurations saturating such a bound.

\subsection{The BPS structure}

\label{sB}

The stationary energy density of the model (\ref{L03}) is
\begin{eqnarray}
\epsilon &=&\frac{1}{2g^{2}}B^{2}+\frac{1}{2g^{2}}(\partial _{i}A_{0}) ^{2}+%
\frac{\lambda ^{2}}{2}(A_{0})^{2}(\vec{n}\cdot \partial _{i}\vec{\phi})^{2}+%
\frac{\lambda ^{2}}{2}Q^{2}  \notag \\[0.2cm]
&&+\frac{1}{2g^{2}}( \partial _{i}\Psi) ^{2}+\frac{\lambda ^{2}}{2}\Psi ^{2}(%
\vec{n}\cdot \partial _{i}\vec{\phi})^{2}+U{(\phi _{n},\Psi )}\text{,}
\label{energy0}
\end{eqnarray}%
where we have used
\begin{equation}
Q^{2}=\frac{1}{2}(D_{i}\vec{\phi}\times D_{j}\vec{\phi})^{2}\text{.}
\end{equation}

Before we establish the field boundary conditions under the requiring for energy density be null when $\left\vert \vec{x}\right\vert \rightarrow \infty $, we set the vacuum condition for the Skyrme field
\begin{equation}
\lim_{\left\vert \vec{x}\right\vert \rightarrow \infty }\vec{\phi}=\hat{n},
\label{bcc00}
\end{equation}%
which provides%
\begin{equation}
\lim_{\left\vert \vec{x}\right\vert \rightarrow \infty }\vec{n}\cdot
\partial _{i}\vec{\phi}=0,\;\text{or}\;\lim_{\left\vert \vec{x}\right\vert
\rightarrow \infty }\partial _{i}\vec{\phi}=\vec{0}.  \label{bcc01}
\end{equation}%
The other boundary conditions arising from energy density (\ref{energy0}) are
\begin{align}
\lim_{\left\vert \vec{x}\right\vert \rightarrow \infty }B &=0,~\
\lim_{\left\vert \vec{x}\right\vert \rightarrow \infty }Q=0,  \label{bcc22}
\\[0.2cm]
\lim_{\left\vert \vec{x}\right\vert \rightarrow \infty }\partial _{i}A_{0}
&=0,~\ \lim_{\left\vert \vec{x}\right\vert \rightarrow \infty }\partial
_{i}\Psi =0,\quad  \label{bcc24} \\[0.2cm]
& \hspace{-0.5cm} \lim_{\left\vert \vec{x}\right\vert \rightarrow \infty }U{%
(\phi _{n},\Psi )}=0.  \label{bcc25}
\end{align}

We still are able to set boundary conditions on the gauge field
and the neutral scalar. Firstly, from (\ref{Q1}) and $\displaystyle%
\lim_{\left \vert\vec{x} \right \vert \rightarrow \infty }Q=0$, we conclude
the potential vector satisfies
\begin{equation}
\color{black}\lim_{\left\vert\vec{x}\right\vert \rightarrow \infty } A_i<\infty. \label{agginf}
\end{equation}
Secondly, from the energy density (\ref{energy0}), the terms $(A_{0})^{2} (%
\vec{n}\cdot \partial_{i}\vec{\phi})^{2}$ and $\Psi^{2} (\vec{n}\cdot
\partial_{i} \vec{\phi})^{2}$ provide the fields $A_0$ and $\Psi$ must
remain finite when $\left\vert\vec{x} \right \vert \rightarrow \infty$.

The total energy is defined by integrating the energy density (\ref{energy0}),
\begin{equation}
E=\int d^{2} x\,\epsilon.
\end{equation}%
Now, with the aim to implement the BPS formalism, we introduce
two auxiliary functions, namely $\Sigma\equiv\Sigma(\phi_{n},\Psi )$ and $%
Z\equiv Z(\phi_{n})$ which we shall determine later. Thus, after some
algebraic manipulations, the total energy reads
\begin{eqnarray}
E &=&\int d^{2}x\left[ \frac{1}{2g^{2}}\left( B\pm \Sigma \right) ^{2}+\frac{%
\lambda ^{2}}{2}\left( Q\mp Z\right) ^{2}\right.  \notag \\[0.08in]
&&+\frac{1}{2g^{2}}\left( \partial _{i}A_{0}\mp \partial _{i}\Psi \right)
^{2}+\frac{\lambda ^{2}}{2}\left( A_{0}\mp \Psi \right) ^{2}(\vec{n}\cdot
\partial _{i}\vec{\phi})^{2}  \notag \\[0.08in]
&&\pm \lambda ^{2}A_{0}\Psi (\vec{n}\cdot \partial _{i}\vec{\phi})^{2}\mp
\frac{1}{g^{2}}B\Sigma -\frac{1}{2g^{2}}\Sigma ^{2}  \notag \\[0.08in]
&&\left. \pm \lambda ^{2}QZ-\frac{\lambda ^{2}}{2}Z^{2}\pm \frac{1}{g^{2}}%
\left( \partial _{i}A_{0}\right) \partial _{i}\Psi +U\right] \text{.\quad
\quad }  \label{en1}
\end{eqnarray}%
By using the expression (\ref{Q1}) and the Gauss law (\ref{gL}), we arrive
at
\begin{eqnarray}
E &=&\int d^{2}x\left[ \frac{1}{2g^{2}}\left( B\pm \Sigma \right) ^{2}+\frac{%
\lambda ^{2}}{2}\left( Q\mp Z\right) ^{2}\right.  \notag \\[0.2cm]
&&+\frac{1}{2g^{2}}\left( \partial _{i}A_{0}\mp \partial _{i}\Psi \right)
^{2}+\frac{\lambda ^{2}}{2}\left( A_{0}\mp \Psi \right) ^{2}(\vec{n}\cdot
\partial _{i}\vec{\phi})^{2}  \notag \\[0.2cm]
&&\pm \lambda ^{2}Z\vec{\phi}\cdot (\partial _{1}\vec{\phi}\times \partial
_{2}\vec{\phi})\pm \frac{1}{g^{2}}\partial _{i}\left( \Psi \partial
_{i}A_{0}\right)  \notag \\[0.2cm]
&&\mp \frac{1}{g^{2}}\epsilon _{ji}(\partial _{j}A_{i})\left( \Sigma -\kappa
\Psi \right) \pm \lambda ^{2}\epsilon _{ij}A_{i}Z(\vec{n}\cdot \partial _{j}%
\vec{\phi})  \notag \\[0.2cm]
&&\left. -\frac{1}{2g^{2}}\Sigma ^{2}-\frac{\lambda ^{2}}{2}Z^{2}+U\right]
\text{.}  \label{enNN}
\end{eqnarray}%
where in the fourth row we have used $B=\epsilon _{ji}\partial_{j}A_{i}$.

{\color{black}At this point, we transform the fourth row of Eq. (\ref{enNN}) in a total derivative. To achieve this, we set}
\begin{equation}
\Sigma \equiv \lambda ^{2}g^{2}W+\kappa \Psi \text{,}
\end{equation}%
being $W\equiv W(\phi _{n})$  and,  thus, the fourth row in (\ref{enNN}) leads to
\begin{equation}
\pm \lambda ^{2}\epsilon _{ij}\left[ (\partial _{j}A_{i})W+A_{i}Z(\vec{n}%
\cdot \partial _{j}\vec{\phi})\right] .
\end{equation}%
It becomes a total derivative by setting
\begin{equation}
\partial _{j}W=Z(\vec{n}\cdot \partial _{j}\vec{\phi})\;\text{ such that }%
\;Z=\frac{\partial W}{\partial \phi _{n}}.
\end{equation}

Therefore, the total energy becomes
\begin{eqnarray}
E &=&\int d^{2}x\left\{ \frac{1}{2g^{2}}\left[ \frac{{}}{{}}\!B\pm \left(
\lambda ^{2}g^{2}W+\kappa \Psi \right) \right] ^{2}\right.  \notag \\[0.2cm]
&&\hspace{-0.5cm}+\frac{\lambda ^{2}}{2}\left[ Q\mp \frac{\partial W}{%
\partial \phi _{n}}\right] ^{2}+\frac{\lambda ^{2}}{2}\left( A_{0}\mp \Psi
\right) ^{2}(\vec{n}\cdot \partial _{i}\vec{\phi})^{2}  \notag \\[0.2cm]
&&\hspace{-0.5cm}+\frac{1}{2g^{2}}\left( \partial _{i}A_{0}\mp \partial
_{i}\Psi \right) ^{2}\pm \lambda ^{2}\left( \frac{\partial W}{\partial \phi
_{n}}\right) \vec{\phi}\cdot (\partial _{1}\vec{\phi}\times \partial _{2}%
\vec{\phi})  \notag \\[0.2cm]
&&\hspace{-0.5cm}\pm \frac{1}{g^{2}}\partial _{i}(\Psi \partial
_{i}A_{0})\mp \lambda ^{2}\epsilon _{ij}\partial _{j}(A_{i}W)  \notag \\%
[0.2cm]
&&\hspace{-0.5cm}\left. -\frac{1}{2g^{2}}\left( \lambda ^{2}g^{2}W+\kappa
\Psi \right) ^{2}-\frac{\lambda ^{2}}{2}\left( \frac{\partial W}{\partial
\phi _{n}}\right) ^{2}+U\right\} \text{,}  \label{enNN0}
\end{eqnarray}

{\color{black}To continue with the implementation of the BPS formalism we require the potential  $U(\phi_{n},\Psi )$ be defined as}
\begin{equation}
U(\phi _{n},\Psi )=\frac{\lambda ^{2}}{2}\left(\frac{\partial W}{\partial
\phi _{n}}\right)^{2}+\frac{\lambda^{4}g^{2}}{2}\left( W+\frac{\kappa }{%
\lambda ^{2}g^{2}}\Psi \right) ^{2}\!,  \label{Vv}
\end{equation}
which is {\color{black}the one able} to generate self-dual configurations. Notably, $W(\phi _{n})$ plays the role of a ``superpotential function", being its structure analogous the one we found {in the literature, e.g., in the
context of} self-gravitating domain walls \cite{Skenderis,DeWolfe, Trigiante}
or scalar field inflation \cite{Townsend,Nardini,Berera,Cicciarella} models.
From henceforth, we shall call the function $W(\phi _{n}) $ {as the
superpotential.}

The function $W(\phi _{n})$ must be constructed (or proposed) such the self-dual potential $U(\phi _{n},\Psi )$ satisfies the vacuum condition expressed in Eq. (\ref{bcc25}). Consequently, a brief analysis of the relation (\ref{Vv}) allows us to impose the following boundary conditions for the superpotential
\begin{equation}
\lim_{\phi _{n}\rightarrow 1}{W}{(\phi _{n})}=0,\ \ \lim_{\phi
_{n}\rightarrow 1}\frac{\partial W}{\partial \phi _{n}}=0,  \label{BcVv}
\end{equation}%
and for the neutral scalar field,
\begin{equation}
\lim_{\left\vert \vec{x}\right\vert \rightarrow \infty }{\Psi }=0.
\label{BcVv0}
\end{equation}

Considering the boundary conditions (\ref{BcVv}) and (\ref{BcVv0}), we observe the contributions of the total derivatives in the fourth row of Eq. (\ref{enNN0}) vanish, i.e.,
\begin{equation}
\int d^{2}x\,\epsilon _{ij}\partial _{j} ( A_{i}W) =0\text{,\ \ } \int
d^{2}x\,\partial _{i}( \Psi \partial _{i}A_{0} ) =0\text{.}  \label{Bc}
\end{equation}%
It is important to emphasize that the second expression in (\ref{Bc}) also would be satisfied by the boundary condition on the electric field given in Eq. (\ref{bcc24}).

Therefore, we consider the total energy written as
\begin{equation}
E=\bar{E}+E_{_{\text{BPS}}}\text{,}  \label{en5}
\end{equation}
where $\bar{E}$ represents the integration composed by the quadratic terms,
\begin{eqnarray}
\bar{E} &=&\int d^{2}x\left\{ \frac{1}{2g^{2}}\left[\frac{}{}\!B\pm ({\kappa
}\Psi +g^{2}\lambda ^{2}W)\right] ^{2}\right.  \notag \\[0.2cm]
&&\quad+\frac{\lambda ^{2}}{2}\left[Q\mp \frac{\partial W}{\partial \phi _{n}%
} \right] ^{2}+\frac{1}{2g^{2}}\left( \partial _{i}A_{0}\mp \partial
_{i}\Psi \right) ^{2}  \notag \\[0.2cm]
&&\left.\quad +\frac{\lambda ^{2}}{2}\left( A_{0}\mp \Psi \right) ^{2}(\vec{n%
} \cdot \partial _{i}\vec{\phi})^{2}\right\} \text{,}  \label{en4}
\end{eqnarray}
and $E_{_{\text{BPS}}}$ defines the energy lower bound,
\begin{equation}
E_{_{\text{BPS}}}=\pm \lambda ^{2}\int d^{2}x\left(\frac{\partial W}{%
\partial \phi _{n}}\right)\vec{\phi}\cdot (\partial _{1}\vec{\phi}\times
\partial _{2}\vec{\phi})\text{.}  \label{en3}
\end{equation}

The total energy (\ref{en5}) satisfy the inequality
\begin{equation}
E\geq E_{_{\text{BPS}}}\text{,}
\end{equation}%
because $\bar{E}\geq 0$. Then, the energy lower bound will be achieved when the fields possess configurations such that $\bar{E}=0$, i.e., the field configurations be solutions of the following set of first-order differential equations:
\begin{equation}
B=\mp g^{2}\lambda ^{2}W\mp {\kappa }\Psi \text{,}  \label{1rd0}
\end{equation}%
\begin{equation}
Q=\pm \frac{\partial W}{\partial \phi _{n}} \text{,}  \label{2rd0}
\end{equation}%
\begin{equation}
\partial _{i}\Psi=\pm \partial _{i}A_{0} \text{, \ }\Psi=\pm A_{0}\text{.}
\label{3rd}
\end{equation}%
They are the so-called self-dual or BPS equations corresponding to the model (\ref{L03}). This set of equations possess a correspondence \cite{Witten, Hlousek} with the BPS equations of an extended supersymmetric model. Indeed, the model (\ref{L03}) is related to the bosonic part of a $\mathcal{N}=2$ SUSY extension \cite{Queiruga}, so that the solutions of the BPS equations are classified as being of the type $1/4$-BPS associated to the existence of a nontrivial phase of the SUSY extended model.

From (\ref{3rd}), we observe that $\Psi=\pm A_{0}$ automatically satisfies both equations, consequently, the self-dual or BPS charged solitons are described by the equations
\begin{equation}
B=\mp g^{2}\lambda ^{2}W - \kappa A_{0} \text{,}  \label{1rd}
\end{equation}%
\begin{equation}
Q=\pm \frac{\partial W}{\partial \phi _{n}} \text{,}  \label{2rd}
\end{equation}
together with the Gauss law (\ref{gL}),
\begin{equation}
\partial _{i}\partial_i A_{0}+\kappa B=g^{2}\lambda ^{2}A_{0}(\vec{n}\cdot
\partial _{i}\vec{\phi})^{2}\text{.}  \label{gL1}
\end{equation}%
{\color{black}In addition, the} boundary condition (\ref{BcVv0}) in the BPS limit implies the scalar potential must satisfy
\begin{equation}
\lim_{\left\vert \vec{x}\right\vert \rightarrow \infty }A_0=0.  \label{BcA0}
\end{equation}

\subsection{\color{black}Equivalence between the BPS and Euler-Lagrange
equations}

\label{sBB}

In this section let us to show that from the BPS equations we recover the
stationary Euler-Lagrange equations provided by the Lagrangian density (\ref%
{L03}), namely, the Amp\`{e}re law (\ref{aL}) and the Skyrme field equation (%
\ref{Fs}).

\subsubsection{\color{black}Recovering the Amp\`{e}re law}

First, we take the partial derivative of the BPS equation (\ref{1rd}) getting%
\begin{equation}
\partial _{i}B=\mp g^{2}\lambda ^{2}\frac{\partial W}{\partial \phi _{n}}%
\partial _{i}\phi _{n}\mp {\kappa }\partial _{i}\Psi \text{,}
\end{equation}%
and now by using the BPS equations (\ref{2rd}) and (\ref{3rd}) we obtain%
\begin{equation}
\partial _{i}B=-g^{2}\lambda ^{2}Q\partial _{i}\phi _{n}-{\kappa }\partial
_{i}A_{0}\text{,}
\end{equation}%
being exactly the stationary Amp\`{e}re law.

\subsubsection{\color{black}Recovering the Skyrme field equation}

With the aim to recover the stationary field equation (\ref{Fs}) for the
Skyrme field we begin rewriting the BPS equation (\ref{2rd}) as
\begin{equation}
Q(D_{j}\vec{\phi})=\pm \frac{\partial W}{\partial \phi _{n}}(D_{j}\vec{\phi})%
\text{,}
\end{equation}%
and applying the quantity $\lambda ^{2}\epsilon _{ij}D_{i}$ in both sides
results%
\begin{eqnarray}
\lambda ^{2}\epsilon _{ij}D_{i}(QD_{j}\vec{\phi}) &=&\pm \lambda
^{2}\epsilon _{ij}D_{i}\left( \frac{\partial W}{\partial \phi _{n}}D_{j}\vec{%
\phi}\right)  \notag \\[0.2cm]
&=&\pm \lambda ^{2}\epsilon _{ij}\frac{\partial ^{2}W}{\partial \phi _{n}^{2}%
}\left( \partial _{i}\phi _{n}\right) D_{j}\vec{\phi}  \notag \\[0.2cm]
&&\pm \frac{\lambda ^{2}}{2}\frac{\partial W}{\partial \phi _{n}}\epsilon
_{ij}\left[ D_{i},D_{j}\right] \vec{\phi}\text{.}
\end{eqnarray}%
Now, we use the following identities%
\begin{equation}
\epsilon _{ij}\left( \partial _{i}\phi _{n}\right) D_{j}\vec{\phi}=-Q(\vec{n}%
\times \vec{\phi})\text{,}
\end{equation}%
\begin{equation}
\epsilon _{ij}\left[ D_{i},D_{j}\right] \vec{\phi}=2B(\vec{n}\times \vec{\phi%
})\text{,}
\end{equation}%
to arrive at%
\begin{eqnarray}
\lambda ^{2}\epsilon _{ij}D_{i}(QD_{j}\vec{\phi}) &=&\mp \lambda ^{2}\frac{%
\partial ^{2}W}{\partial \phi _{n}^{2}}Q(\vec{n}\times \vec{\phi})  \notag \\%
[0.2cm]
&&\pm \lambda ^{2}\frac{\partial W}{\partial \phi _{n}}B(\vec{n}\times \vec{%
\phi})\text{.}
\end{eqnarray}%
Inserting the BPS equations (\ref{1rd}) and (\ref{2rd}), just in the right
side, we obtain%
\begin{align}
\lambda ^{2}\epsilon _{ij}D_{i}(QD_{j}\vec{\phi})& =-\lambda ^{2}\frac{%
\partial ^{2}W}{\partial \phi _{n}^{2}}\frac{\partial W}{\partial \phi _{n}}(%
\vec{n}\times \vec{\phi})  \notag \\[0.2cm]
& -\lambda ^{2}\frac{\partial W}{\partial \phi _{n}}\left( g^{2}\lambda
^{2}W+{\kappa }\Psi \right) (\vec{n}\times \vec{\phi})\text{.}
\end{align}%
{The right-hand can be write in terms of the derivative of the potential (%
\ref{Vv}) with respect to $\phi_n$, thus, }
\begin{equation}
\lambda ^{2}\epsilon _{ij}D_{i}(QD_{j}\vec{\phi})=-\frac{\partial U}{%
\partial \phi _{n}}(\vec{n}\times \vec{\phi})\text{.}
\end{equation}%
It is the Skyrme field equation (\ref{Fs}) in Bogomol'nyi limit. Therefore,
we have verified that the BPS equations imply in both stationary field
equations, Amp\`{e}re law and the Skyrme field equation.

\section{Rotationally Symmetric Skyrmions \label{bpssolutions}}

We investigate solitons rotationally symmetric saturating the energy lower bound (\ref{en3}). Henceforth, without loss of generality, we set $\vec{n}=(0,0,1)$ such that $\phi _{n}=\phi_{3}$ and set the usual \textit{ansatz} for the Skyrme field \cite{gladikowski},
\begin{equation}
\vec{\phi}\left( r,\theta \right) =\left(
\begin{array}{c}
\sin f(r)\cos N\theta \\
\sin f(r)\sin N\theta \\
\cos f(r)%
\end{array}%
\right) \text{,}  \label{Antz}
\end{equation}%
where $r$ and $\theta $ are polar coordinates, $N=\deg [\vec{\phi}]$ is the
winding number introduced in (\ref{wn}) and  $f(r)$ a regular function satisfying boundary conditions
\begin{equation}
f(0)=\pi \text{,}~\ \ \lim_{r\rightarrow \infty }f(r)=0\text{.}
\label{bccr1}
\end{equation}
The representation (\ref{Antz}) is a two-dimensional version of the hedgehog ansatz used in the three-dimensional Skyrme model \cite{Weigel}.

For the gauge field $A_{\mu}$, we consider the {ansatz}
\begin{equation}
A_{i}=-\epsilon _{ij}{x}_{j}\frac{Na(r)}{r^2} \text{, \ }A_{0}=\omega(r)
\text{,}
\end{equation}
where the profile functions $a(r)$ and $\omega(r)$ are well behaved
functions satisfying the boundary conditions,
\begin{equation}
a(0)=0\text{, \ }\lim_{r\rightarrow \infty }a(r)=a_{\infty }\text{,}
\label{bccr2}
\end{equation}
\begin{equation}
\omega(0)=\omega_0\text{, \ }\lim_{r\rightarrow \infty }\omega(r)=0\text{, \
} \lim_{r\rightarrow\infty }\frac{d\omega }{dr}=0\text{,}  \label{bccr3}
\end{equation}
where $a_{\infty }$ and $\omega_0$ {\color{black}are finite constants.}

By convenience, we now introduce the field redefinition \cite{adam2} as follows:
\begin{equation}
\phi _{3}=\cos f\equiv 1-2h\text{,}
\end{equation}
with the field $h=h(r)$ obeying
\begin{equation}
{\color{black}h(0) =1}\text{, \ \ }\lim_{r\rightarrow \infty }h(r) =0.
\label{bccr4}
\end{equation}

The boundary conditions for the superpotential $W(h)$ are
\begin{equation}
\lim_{r\rightarrow 0}W(h)=W_0\text{, \ }\lim_{r\rightarrow\infty}W(h) =0%
\text{, \ }\lim_{r\rightarrow \infty}\frac{dW}{dh} =0\text{,}  \label{c03}
\end{equation}
where $W_0$ is a positive finite constant and the two last conditions are
consequence of Eq. (\ref{BcVv}).

{\color{black}The magnetic field and the electric field result be
\begin{equation}
B=\frac{N}{r}\frac{da}{dr}\text{,}\ \ \ \ E_{r}=-\frac{d\omega }{dr}\text{.}
\label{MgF}
\end{equation}
whereas the energy lower bound (\ref{en3}) or BPS bound
becomes
\begin{equation}
E\geq E_{_{\text{BPS}}}=\pm2\pi \lambda ^{2}  N  W_0 \text{.}
\label{en7}
\end{equation}
where $W_0$ is a constant defined in Eq. (\ref{c03}).}

Under the {ansatz}, the BPS equations become
\begin{equation}
\frac{N}{r}\frac{da}{dr}+\lambda ^{2}g^{2}W+\kappa \omega =0\text{,}
\label{bc01}
\end{equation}%
\begin{equation}
\frac{4N}{r}\left( 1+a\right) \frac{dh}{dr}+\frac{dW}{dh}=0\text{,}
\label{bc02}
\end{equation}
while the Gauss law (\ref{gL1}) gives
\begin{equation}
\frac{d^{2}\omega }{dr^{2}}+\frac{1}{r}\frac{d\omega }{dr}+\frac{\kappa N}{r}%
\frac{da}{dr}=4\lambda ^{2}g^{2}\omega \left( \frac{dh}{dr}\right) ^{2}\text{%
.}  \label{Glaw}
\end{equation}%
Note that in the BPS equations (\ref{bc01}) and (\ref{bc02}), without a loss of generality, we have chosen the upper sign. Such a assumption will be considered in the remaining of the manuscript.

Similarly, we write the self-dual potential
\begin{equation}
U(h,\omega )=\frac{\lambda ^{2}}{8}\left(\frac{dW}{dh}\right)^2+\frac{%
\lambda ^{4}g^{2}}{2}\left( W+\frac{\kappa }{\lambda ^{2}g^{2}}\omega
\right) ^{2}\text{,}  \label{Vsfd}
\end{equation}
and by using the BPS equation the energy density becomes
\begin{equation}
\epsilon _{_{\text{BPS}}}=\frac{B^{2}}{g^{2}}+\frac{1}{g^{2}}\! \left( \frac{%
d\omega}{dr}\right) ^{2}\!\! +4\lambda ^{2}\omega ^{2}\! \left( \frac{dh}{dr}
\right) ^{2}\!\! +\frac{\lambda ^{2}}{4}\! \left(\frac{dW}{dh}\right)^2
\text{,}
\end{equation}%
we call it the BPS energy density.

{\color{black}In the next subsections we study the behavior at origin and when $r\rightarrow\infty$ of the self-dual profiles by solving the BPS equations (\ref{bc01}), (\ref{bc02}) and the Gauss law (\ref{Glaw}).}

\subsection{\color{black}Behavior of the profiles at origin}

We perform the analysis around the origin ($r=0$) by considering the following boundary conditions
\begin{equation}
h(0) =1,\;\;a(0) =0,\;\;\omega(0) =\omega_{0},\;\;\lim_{r\rightarrow 0}W(h)
=W_{0} \text{,}
\end{equation}%
where the superpotential $W(h)$ is considered to be a well-behaved function
with $W_{0}$ being a positive finite quantity. Hence, we find the following
behavior for the field profiles:%
\begin{align}
h(r) &\approx\color{black} 1-\frac{\left( W_{h}\right) _{h=0}}{8N}r^{2}+ \frac{\left(W_{h}\right) _{h=0}\left( W_{hh}\right)_{h=0}}{128N^{2}} r^{4}\text{,} \\[0.2cm]
a(r) &\approx \color{black}-C_{0}r^{2}+\frac{g^{2}\lambda ^{2}\left( W_{h}\right)_{h=0}^{2}-4\kappa ^{2}N^{2}C_{0}}{32N^{2}}r^{4}\text{,} \\[0.2cm]
\omega (r) &\approx \omega _{0}+\frac{\kappa NC_{0} }{2}r^{2}\notag\\[0.2cm]
&\color{black}\quad-\frac{{\kappa} g^{2}\lambda ^{2} \left( W_{h}\right)
_{h=0}^{2}-4\kappa ^{3}N^{2}C_{0}}{128N}r^{4}\text{,}
\end{align}
{\color{black}where $W_{h}= {d W}/{dh}$, $W_{hh}= {d^2 W}/{dh^2}$, and we
have defined the constant}
\begin{equation}
C_{0}=\frac{g^{2}\lambda ^{2}W_{0}+\kappa \omega _{0}}{2N}\text{.}
\end{equation}

The behavior for magnetic and electric field near to the origin are
\begin{align}
B(r)& \approx\color{black} -2NC_{0}+\frac{g^{2}\lambda ^{2}\left( W_{h}\right)_{h=0}^{2}-4N^{2}\kappa ^{2}C_{0}}{8N}r^{2}\text{,} \\[0.3cm]
E_{r}(r)& \approx \color{black}-{\kappa }NC_{0}r+\frac{\kappa g^{2}\lambda ^{2}\left(
W_{h}\right) _{h=0}^{2}-4\kappa ^{3}N^{2}C_{0}}{32N}r^{3}\text{,}
\end{align}%
respectively, while the BPS energy density gives%
\begin{eqnarray}
\epsilon _{_{\text{BPS}}} &\approx &\frac{4N^{2}C_{0}^{2}}{g^{2}}+\frac{%
\lambda ^{2}}{4}\left( W_{h}\right) _{h=0}+\left[ \frac{3\kappa
^{2}N^{2}C_{0}^{2}}{g^{2}}\right.  \notag \\[0.3cm]
&&\color{black}+\frac{\lambda ^{2}\omega _{0}^{2}\left( W_{h}\right) _{h=0}^{2}}{4N^{2}}-\frac{\lambda ^{2}C_{0}\left( W_{h}\right) _{h=0}^{2}}{2}  \notag \\[0.3cm]
&&\color{black}\left. -\frac{1}{16}\frac{\lambda ^{2}\left( W_{h}\right) _{h=0}^{2}\left(W_{hh}\right) _{h=0}}{N}\right] r^{2}\text{.}
\end{eqnarray}
{\color{black}It is verified that the amplitude of the BPS energy density at the origin increase in accordance with the growth of the electromagnetic coupling $g$.}

\subsection{Behavior of the profiles for large values of $r$}

The behavior of the fields when $r\rightarrow \infty$ is performed by taking the following boundary conditions,
\begin{equation}
h(\infty) =0,\;\;a(\infty) =a_{\infty},\;\; \omega(\infty)=0,\;\;
\lim_{r\rightarrow\infty} W(h) =0\text{.}  \label{asyBC}
\end{equation}%
Besides that, we consider a superpotential $W(h)$  behaving as
\begin{equation}
W(h) \approx\frac{h^{\sigma }}{\lambda ^{2}}\text{,}  \label{Whv}
\end{equation}%
with the parameter $\sigma \geq 2$.  Under the boundary conditions (\ref%
{asyBC}), the asymptotic analysis leads us to two type {\color{black}of Skyrme field profiles: {\color{black}(i) for} $\sigma =2$ we have found soliton solutions whose tail decays following an exponential-law type $e^{-\alpha r^2}$ ($\alpha>0$); {\color{black}(ii) for $\sigma >2$} the profiles of the Skyrme field have a power-law decay $r^{-\beta}$ ($\beta>0$). At same time, {\color{black}for $\sigma\geq 2$} the gauge field profiles $a(r)$ and $\omega(r)$ decay following an exponential-law type $e^{-m r}\left( m >0\right) $. }

\subsubsection{Behavior of the profiles for $\protect\sigma =2$}

{\color{black}{\color{black} We take a superpotential whose behavior is}}
\begin{equation}
W(h)\approx\frac{h^{2}}{\lambda ^{2}}\text{,}  \label{Wh2a}
\end{equation}
{\color{black}{\color{black} such that} the field profiles possess the following asymptotic behavior }
\begin{align}
h(r) &\color{black}\approx C^{(h)}_{\infty}e^{-\Lambda r^{2}} \text{,}  \label{hsig2} \\[0.2cm]
a(r) &\approx a_{\infty }\color{black}-C_{\infty} \sqrt{r}e^{-\kappa r} \text{,}  \label{aR} \\[0.2cm]
\omega (r) &\approx\color{black}  -C_{\infty }\frac{N}{ \sqrt{r}}%
e^{-\kappa r} \text{,}  \label{aW}
\end{align}
where $C^{(h)}_{\infty}$ and $C_{\infty }$ are arbitrary constants and the quantity $%
\Lambda $ has been defined as
\begin{equation}
\Lambda =\frac{1}{4N\lambda ^{2}\left( 1+a_{\infty }\right) }\text{.} \label{expsq}
\end{equation}

{\color{black}In addition,  the magnetic and electric fields for large values of $r$ behave as} {\color{black}
\begin{align}
B(r) &\approx  C_{\infty }\frac{N\kappa}{\sqrt{r}}e^{-\kappa r}
\text{,}  \label{bhB} \\[0.2cm]
E_{r}(r) &\approx  - C_{\infty }\frac{N\kappa}{\sqrt{r}}e^{-\kappa
r} \text{,}  \label{bhEr}
\end{align}
respectively. We must highlight the gauge field behavior (including the electric and magnetic fields) resembles the one of the Abrikosov-Nielsen-Olesen vortices arising in Abelian Higgs models \cite{Nielsen,Jackiw,Paul}.  Remembering the Maxwell-Higgs electrodynamics is the relativistic counterpart of the Ginzburg-Landau theory of superconductivity and, the BPS limit separates the phases  describing the Type-I and Type-II superconductivity. {\color{black}Furthermore, the behavior shows explicitly the Chern-Simons coupling constant $\kappa$ plays the role of the effective mass of the gauge field.}}

\subsubsection{\color{black}Behavior of the profiles for $\protect\sigma >2$}

{\color{black}{\color{black}In this case, we consider a superpotential that obeys}
\begin{equation}
W(h) \approx\frac{h^{\sigma}}{\lambda ^{2}}\text{.}  \label{Whvm2a}
\end{equation}
Hence, when $r\rightarrow\infty$, the field profiles have the following asymptotic {\color{black}behavior:}}
\begin{align}
h(r) &\approx \color{black}\left(\frac{\mathcal{C}^{(h)}_\infty}{r^{2}}\right) ^{1/\left( \sigma-2\right) } \text{,}  \label{hsigS} \\[0.3cm]
a(r) &\approx a_{\infty } \color{black}- \mathcal{C}_{\infty } \sqrt{r}e^{-\kappa r} \text{,} \\[0.3cm]
\omega (r) &\approx  \color{black}-\mathcal{C}_{\infty }\frac{N}{\sqrt{r}} e^{-\kappa r} \text{,}
\end{align}
where $\mathcal{C}_{\infty }$ is an arbitrary constant {\color{black}and $\mathcal{C}^{(h)}_\infty$} defined as
\begin{equation}
\color{black}\mathcal{C}^{(h)}_\infty=\frac{8N\lambda^2\left( a_{\infty }+1\right) }{\sigma \left( \sigma -2\right) }\text{.}
\end{equation}

{\color{black} We note the profiles of the Skyrme field follow a power-law decay contrasting the behavior for $\sigma=2$ given in Eq. (\ref{hsig2}). {\color{black} Field profiles following a power-law decay for large distances} are named \textit{delocalized}. This type of solutions have appeared in the study of two-component superconductors \cite{EBabev}, diamagnetic vortices \cite{xxzz} and in some $k$-generalized Abelian Higgs models \cite{Casana03}. On the other hand, the gauge field profiles $a(r)$ and $\omega(r)$ remain localized because they follow the same behavior of the ones analyzed in the previous case $\sigma=2$ {\color{black}and}, consequently, {\color{black}the magnetic and electric field behaviors are given by Eqs. (\ref{bhB}) and (\ref{bhEr}), respectively.}}

{\color{black}
\section{Numerical solutions\label{numerical}}}

\subsection{\color{black}Numerical solutions for $\sigma=2$}

\begin{figure}[H]
\centering\includegraphics[width=8.3cm]{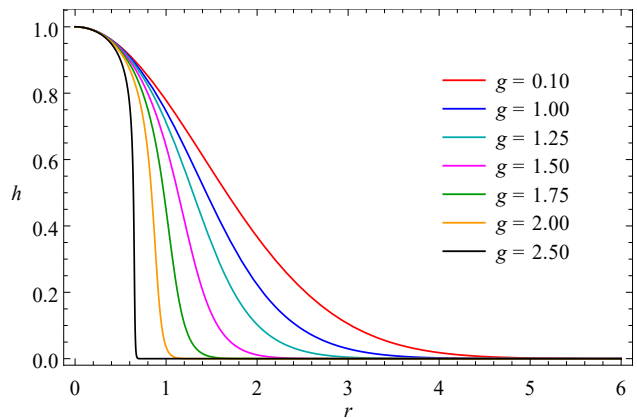}
\caption{Skyrme field profiles $h(r)$.} \label{Fig1}
\end{figure}

\begin{figure}[H]
\centering\includegraphics[width=8.35cm]{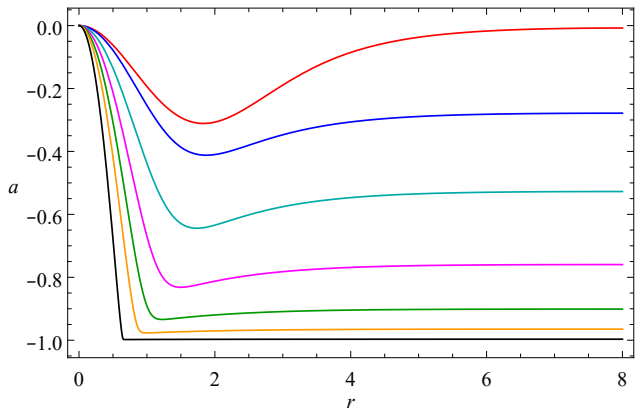}  \caption{{\color{black}Vector potential} profiles $a(r)$. The profile for $g=0.1$ (red line) has been rescaled by multiplying by $100$. Conventions as in Fig. \ref{Fig1}.} \label{Fig2}
\end{figure}

\begin{figure}[H]
\centering\includegraphics[width=8.3cm]{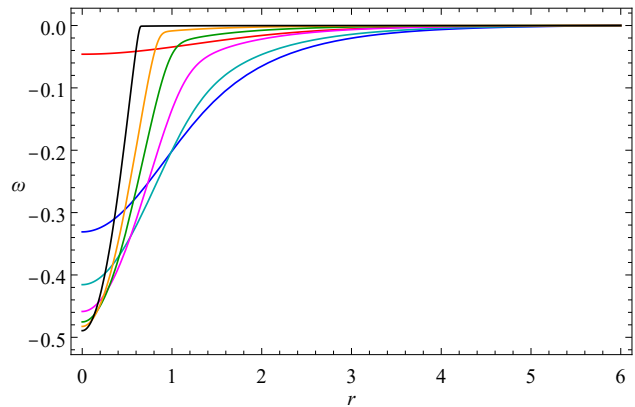}
\caption{{\color{black}Scalar potential} $\protect\omega(r)$. The profile for $g=0.1$ (red line) has been rescaled by multiplying by $10$. Conventions as in Fig. \ref{Fig1}.} \label{Fig3}
\end{figure}

\begin{figure}[H]
\centering\includegraphics[width=8.5cm]{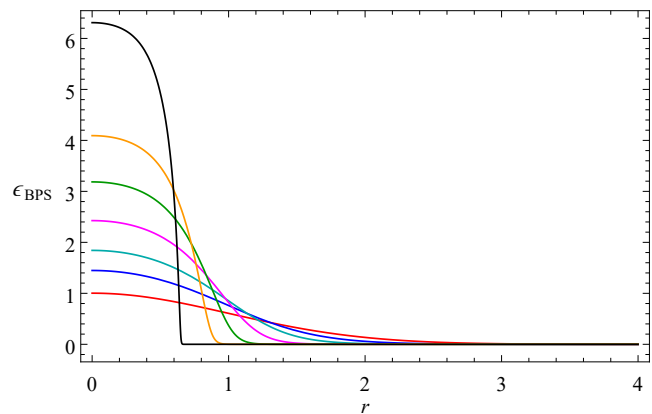}
\caption{BPS energy density $\epsilon_{_{\text{BPS}}}(r)$. Conventions as in Fig. \ref{Fig1}.} \label{Fig4}
\end{figure}

\begin{figure}[t]
\centering\includegraphics[width=8.6cm]{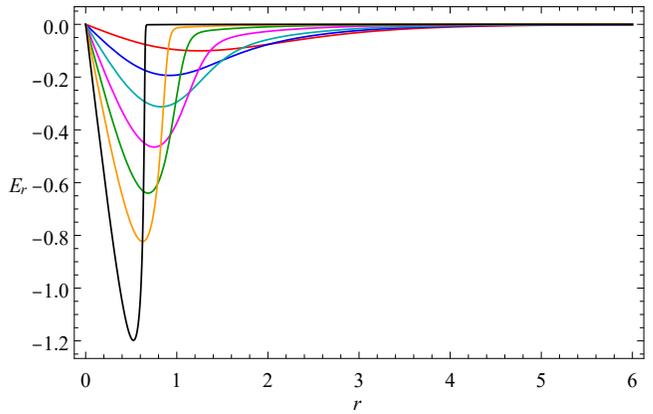}
\caption{Electric field $E_{r}(r)$. The profile for $g=0.1$ (black dotted line) has been rescaled by multiplying by $50$. Conventions as in Fig. \ref{Fig1}.} \label{Fig5}
\end{figure}

\begin{figure}[H]
\centering\includegraphics[width=8.3cm]{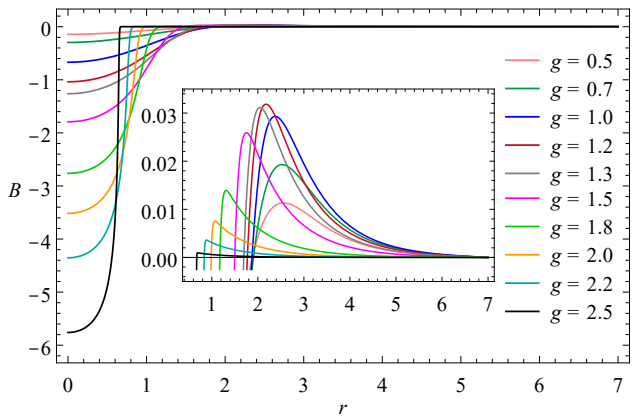}
\caption{Magnetic field $B(r)$.} \label{Fig6}
\end{figure}

{\color{black}Our first numerical analysis is devoted to solve the BPS equations  (\ref{bc01}), (\ref{bc02}) and the Gauss law  (\ref{Glaw}) by considering the superpotential
\begin{equation}
W(h)=\frac{h^{2}}{\lambda ^{2}}\text{.}  \label{Wh2}
\end{equation}
Further, we set $N=1$, $\lambda=1$, $\kappa=1$, and running the electromagnetic coupling constant $g$. The resulting solutions} are shown in Figs. \ref{Fig1}-\ref{Fig6}.

The profile functions $h(r)$ characterizing the Skyrme field are plotted in
Fig. \ref{Fig1}. Note that for {\color{black}increasing values of $g$ the profiles become more localized around the origin. {\color{black}Also,} for sufficiently large values of $g$ (in our analysis, $g\gtrsim 2.5$), the profiles rapidly attain the vacuum value acquiring a structure seeming the one of a compacton   (soliton of finite extent having its exact vacuum value outside of the compact region \cite{Gisiger}).  The arising of the compactonlike structure is consistent with the behavior (\ref{hsig2}) of the Skyrme field profile which yielding a fast exponential decay  in the strong coupling limit of $g$, i.e., the parameter $\Lambda \rightarrow\infty$ as a consequence of $a_{\infty} \rightarrow -1$  (see Fig. \ref{Fig2}).}

{\color{black}Figure \ref{Fig2} depicts the vector potential profiles $a(r)$}. {\color{black}Unlike of the case of the uncharged BPS solitons solutions approached in \cite {adam2}, here the vector potential profiles present a inverted ringlike shape (in our analysis such a feature is better seen in the interval  $0<g<2$) which goes vanishing for sufficiently large values of $g$ when the vector potential achieves a constant vacuum value $a_{\infty}\rightarrow -1$. As previously commented,  when the vacuum value $a_{\infty} \rightarrow-1$ the format of the soliton profiles become compactonlike structures. It is worth emphasize that the arising of ringlike structures is associated to the presence of Chern-Simons term. {\color{black}In order to better visualize} the ringlike effect for weak coupling, the profile for $g=0.1$ (red line) was rescaled by multiplying it by $100$.}

The profiles for the {\color{black}scalar potential $\omega(r)$ and BPS} energy density $\epsilon_{_{\text{BPS}}}(r)$ are shown in Figs. \ref{Fig3} and \ref{Fig4}, respectively.  {\color{black}In both cases the amplitude at origin (in absolute value)   increases  with the growing of $g$. It is again observed the profiles  for sufficiently large values of $g$ present the compactlike format.}

{\color{black}The profiles of the electric field $E_{r}(r)$ are present in Fig. \ref{Fig5}. The profiles are  rings whose maximum amplitude is localized  closer to the origin as the coupling constant $g$ grows besides, they  gain the compactonlike form. {\color{black}Furthermore,}  we have do a rescaling (multiplying by 50) the profile for $g=0.1$ (red line) in order to the ring format becomes more visible. The numerical solutions tell us the electric field is negative for all values of $r$ and $g$.}

{\color{black}Figure \ref{Fig6} depicts the profiles of the magnetic field $B(r)$ for a set of values of the coupling constant $g$. At first sight, we observe whenever $g$ increases, the absolute value of the amplitude in $r=0$ also increases besides the profiles become more localized around the origin acquiring a compactonlike format. However, for  sufficiently large values of $r$,} {\color{black}a close zoom on the profiles (see insertion in Fig. \ref{Fig6}) revels a flipping (signal inversion) of the magnetic field which directly implies in a localized magnetic flux inversion}. {\color{black} Such a flipping of the magnetic field becomes clearer by seeing Eqs. (\ref{bhB}) and (\ref{bhEr}) given the behavior of the magnetic and electric fields, they telling us that for large values of $r$ the fields have opposite signals. Thus, being the electric field always negative, for large distances the magnetic field will be positive. In our analysis the maximum amplitude of the {inversion} grows in the interval of $0<g\leq1.2$, thereafter, decreases continuously for $g>1.2$  until it disappears {for  sufficiently large values} of the coupling constant $g$. We can highlight such a localized magnetic flux inversion in the present BPS model is a genuine effect due to presence of the Maxwell-Chern-Simons term, once that such a effect is absent in others  gauged  BPS baby Skyrme models in which  the Maxwell's contribution \cite{adam2} or  the Chern-Simons contribution \cite{Casana01} have been analyzed individually.}

{The flipping of the magnetic field is a peculiar phenomenon which also has {been reported on some other} $(2+1)$-dimensional systems. For example, {such a behavior arises} in the study of two-component superconductors whose  fractional vortices present a delocalized magnetic field \cite{EBabev}. It also occurs in some Lorentz-violating Maxwell-Higgs electrodynamics  \cite{Casana04,Casana05,Casana07} or in the context of Lorentz-violating gauged $O(3)$ $\sigma$-model \cite{Casana06}.}

\subsection{\color{black}Numerical solutions for $\sigma>2$}

\begin{figure}[H]
\centering\includegraphics[width=8.3cm]{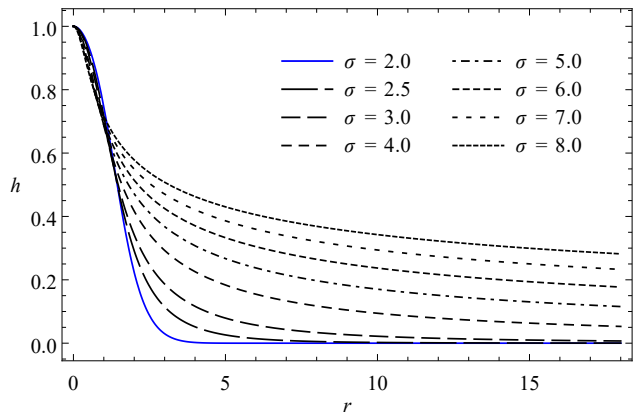}
\caption{Skyrme field profiles $h(r)$ for {different} values of $\protect\sigma$ in superpotential (\protect\ref{Whvm2}).} \label{Fig7}
\end{figure}

\begin{figure}[H]
\centering\includegraphics[width=8.4cm]{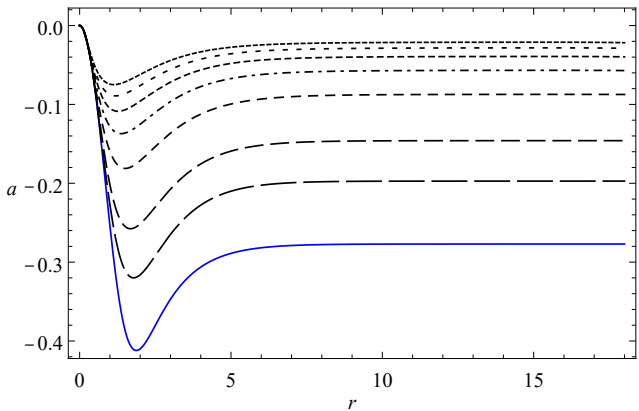}
\caption{Vector potential profiles $a(r)$. Conventions as in Fig.\ref{Fig7}.}  \label{Fig8}
\end{figure}

\begin{figure}[H]
\centering\includegraphics[width=8.6cm]{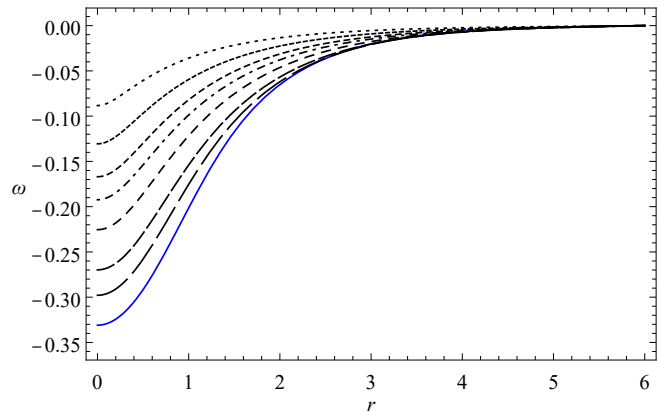}
\caption{Scalar potential $\omega(r)$. Conventions as in Fig. \ref{Fig7}.} \label{Fig9}
\end{figure}

\begin{figure}[t]
\centering\includegraphics[width=8.7cm]{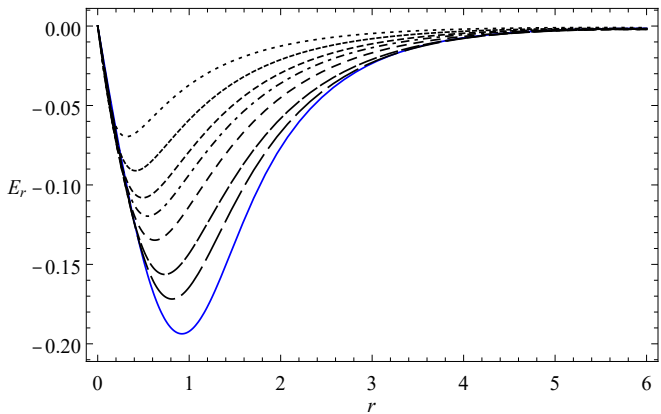}
\caption{Electric field $E_{r}(r)$. Conventions as in Fig. \ref{Fig7}.} \label{Fig10}
\end{figure}

\begin{figure}[]
\centering\includegraphics[width=8.6cm]{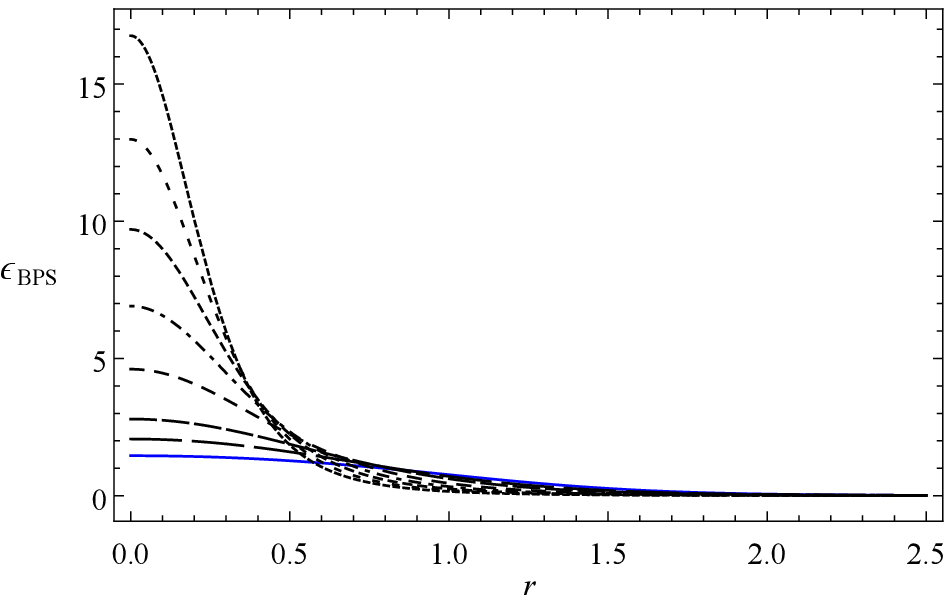}
\caption{BPS density energy $\epsilon_{_{\text{BPS}}}(r)$. Conventions as in Fig. \ref{Fig7}.} \label{Fig11}
\end{figure}

\begin{figure}[]
\centering\includegraphics[width=8.4cm]{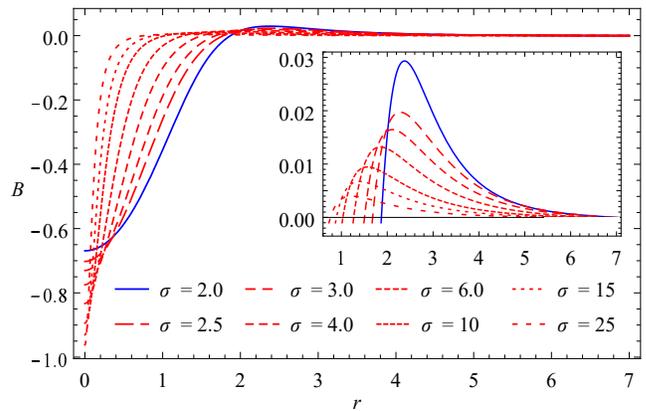}
\caption{Magnetic field  $B(r)$ for {\color{black}other} values of $\sigma$ in  superpotential (\protect\ref{Whvm2}).} \label{Fig12}
\end{figure}

{\color{black}Our second numerical analysis  is performed by considering the superpotential
\begin{equation}
W(h) =\frac{h^{\sigma}}{\lambda ^{2}}\text{,}  \label{Whvm2}
\end{equation}
to solve the set of equations  (\ref{bc01}), (\ref{bc02}), (\ref{Glaw}) for different values of the parameter $\sigma$ and fixing $N=1$, $\lambda=1$, $\kappa=1$, $g=1$. The numerical profiles are shown in Figs. \ref{Fig7}-\ref{Fig12}.}

{\color{black}
Figure \ref{Fig7} shows clearly that for $\sigma >2$ the Skyrme field profile $h(r)$   decay more slowly to its vacuum value whenever  $\sigma$ increases, in according with the power-law given in Eq. (\ref{hsigS}).}

{\color{black}
The general characteristics for profiles of the vector potential $a(r)$, scalar potential $\omega (r)$, electric field $E_{r}(r)$, self-dual energy density $\epsilon_{_{\text{BPS}}}(r)$ and magnetic field $B(r)$ are similar to the ones already described in previous subsection, but becomes more and more localized near to at origin with the growth of the parameter $\sigma $, see Figs. \ref{Fig8}-\ref{Fig12}. {\color{black}Nevertheless, it is worthwhile some comments about the vector potential and magnetic field. In Fig. \ref{Fig8}, we note the ringlike structures of the vector potential profiles are vanishing as increasing of $\sigma$ whereas $a(r)$ tends to zero. As consequence of such a behavior,  the flip of the magnetic field  also is present and its maximum amplitude diminishes with the increasing of the $\sigma$, see Fig. \ref{Fig12}.}}

\subsection{Magnetic flux and electric charge}

{\color{black}The numerical results presented in the cases of $\sigma=2$  allow  us to analyze important results about the magnetic flux and the total electric charge.}

{\color{black}The total magnetic flux is
\begin{equation}
\Phi =2\pi \int_0^\infty rdr\,B\equiv 2\pi Na_{\infty },  \label{flux}
\end{equation}
where we have considered the boundary conditions established in Eq. (\ref{bccr2}), because of the boundary condition (\ref{agginf}) the parameter $a_{\infty}$ is a finite real constant. Therefore, the magnetic flux is in general a  nonquantized quantity (in the topological sense), unlike the one belonging to the Chern-Simons Abelian Higgs models \cite{Khare,Vega}. However, recent investigations have been shown it is possible to obtain quantized magnetic flux in some Skyrme models \cite{adam6,samoilenka4, samoilenka2}.}

{\color{black}From expression (\ref{QQq}), the total electric charge yields
\begin{equation}
\mathcal{Q}_{\text{em}}=-\frac{2\pi\kappa {N}}{g^{2}}a_{\infty }\text{,}  \label{cf2}
\end{equation}
showing it is  nonquantized, too.}

\begin{figure}[H]
\centering\includegraphics[width=8.3cm]{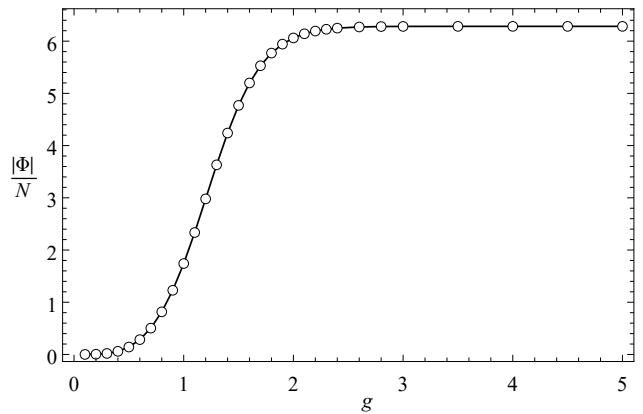}
\caption{The magnetic flux $\left\vert \Phi \right\vert$ in units of $N$ as a function of the gauge coupling $g$ for the superpotential (\protect\ref{Wh2}), fixing $\protect\lambda=1$ and $\protect\kappa=1$.} \label{Fig13}
\end{figure}

{\color{black}We observe in the description of the Fig. {\ref{Fig2}} that for a sufficiently strong coupling $g$ the vacuum value of the potential vector $a_{\infty} \rightarrow -1$, thus, the magnetic flux (\ref{flux}) becomes quantized in this limit, such as it is shown in  Fig.  \ref{Fig13}. This effective quantization implies that the total electric charge (\ref{cf2}) also becomes quantized in such regime.}

\begin{figure}[t]
\centering\includegraphics[width=8.6cm]{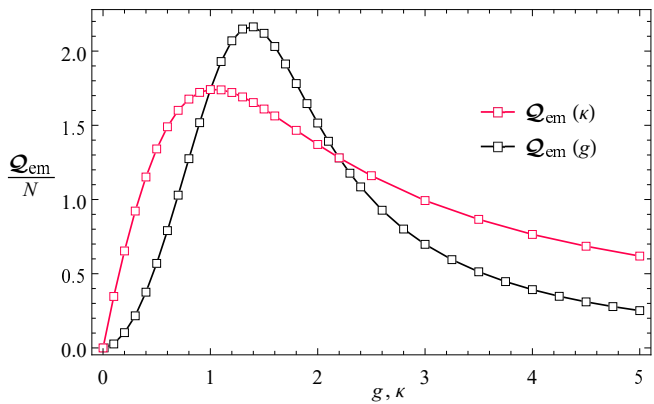}
\caption{{\color{black}The total electric charge in units of $N$ as a function of both the constant electromagnetic coupling $\mathcal{Q}_{\text{em}}(g)$ and Chern-Simons coupling $\mathcal{Q}_{\text{em}}(\kappa)$ for the superpotential (\protect\ref{Wh2}) with $\protect\lambda=1$.}} \label{Fig14}
\end{figure}

{\color{black}{\color{black}Figure} \ref{Fig14} we depict the behavior of the total electric charge as a function of the coupling constants $g$ [$\mathcal{Q}_{\text{em}} (g)$ with $\kappa$ fixed] and $\kappa$ [$\mathcal{Q}_{\text{em}} (\kappa)$ with $g$ fixed]  {\color{black}by adopting} the superpotential (\ref{Wh2}) with $\lambda=1$. For the first analysis, it is fixed $\kappa=1$ (black line-squared), we note {\color{black}the total electric charge $\mathcal{Q}_{\text{em}} (g)$} increases in accordance with $g$ until it attains its maximum value at $g_{\text{max}}\simeq1.387$, {\color{black}after} the electric charge diminishes when $g$  increases continuously, i.e.,  {\color{black}$\mathcal{Q}_{\text{em}} (g) \sim g^{-2}$ for $g\gg g_{\text{max}}$,} compatible with Eq. (\ref{cf2}).}

\begin{figure}[]
	\centering\includegraphics[width=8.5cm]{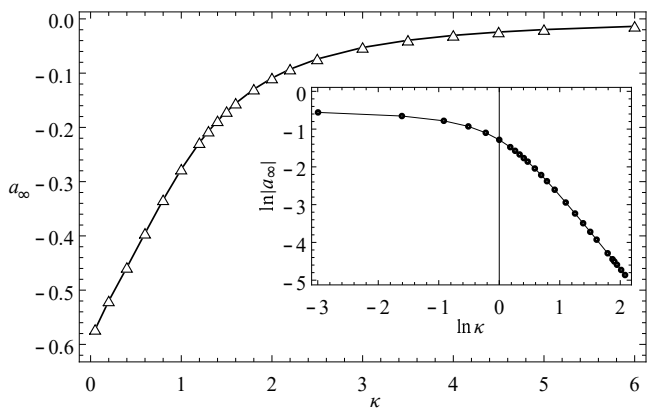}
	\caption{The gauge vacuum value $a_{\infty}$ as a function of the Chern-Simons coupling $\protect\kappa$ by assuming the superpotential (\protect\ref{Wh2}) with  {\color{black}$g=1$, $\protect\lambda=1$ and $N=1$.  The insertion shows} the logarithm of the absolute value of $a_{\infty}$ as function of the logarithm of $\kappa$.}
	\label{Fig15}
\end{figure}

{\color{black}The second analysis presented in Fig. \ref{Fig14} is performed by considering the gauge coupling constant fixed, $g=1$ (red line-squared).} We observe the {\color{black}the total electric charge $\mathcal{Q}_{\text{em}} (\kappa)$} grows as $\kappa$ and reaches its maximum value at $\kappa_{\text{max}} \simeq 1.045$, from then on, gradually lessens with the continuous growth of the {\color{black}Chern-Simons coupling}. {\color{black} Although that behavior is not {\color{black}directly} explained by Eq. (\ref {cf2}), we find numerically that for the interval $0<\kappa<\kappa_{\text{max}} $ the vacuum value $a_{\infty}$ {\color{black} {\color{black} seems to grow linearly with $\kappa$}, see Fig. \ref{Fig15}. Already for $\kappa>\kappa_{ \text{max}}$, the values of $a_{\infty}$ {\color{black}increase slower} than growth of coupling $\kappa$; this  behavior can be better understood by analyzing the insertion into Fig. \ref{Fig15}, where we have $\ln\left\vert a_{\infty} \right\vert$ as function of $\ln\kappa$: we note that $\left\vert a_{\infty} \right\vert\sim\kappa^{-2}$ for $\kappa\gg \kappa_{ \text{max}}$.} Such an offbeat behavior is similar to the one of topological vortices obtained in the Chern-Simons $O(3)$ $\sigma$-model discussed in \cite{Gladikowski2}.}

{\color{black}Now let us briefly comment about the magnetic flux and total electric charge for the case $\sigma>2$. In Fig. \ref{Fig8}, {\color{black}we note that for a fixed value of the gauge coupling $g$, the vacuum value $a_{\infty}$  goes to zero continuously as $\sigma$ grows, i.e., $\displaystyle{\lim_{\sigma \rightarrow \infty}  a_{\infty} = 0}$.  Consequently, the magnetic flux  and the total electric charge {become null}.} An analogous result was obtained in the  generalized Chern-Simons baby Skyrme model \cite{Casana01}.}

\section{Conclusions and remarks \label{conclusion}}

{\color{black}We have shown the existence of BPS charged configurations in a gauged baby Skyrme model $(\ref{L03})$ whose gauge field is governed by the Maxwell-Chern-Simons action. The BPS model $(\ref{L03})$ is constructed by introducing a scalar field $\Psi$ into the model (\ref{L0}) which couples adequately to the Skyrme field $\vec{\phi}$ but it does not couples to the gauge field.   The successful implementation of the BPS technique allows to obtain the energy lower bound (is related to the topological charge of the Skyrme field) and hence the self-dual or BPS equations  whose solutions saturate this bound. We point out the introduction of a superpotential function determining the self-dual potential is the important step in the successful implementation of the BPS technique.  Such a superpotential is considered to be a well-behaved function in the whole target space and plays the important role by defining the BPS configurations.} {\color{black}It is worth mentioning the model $(\ref{L03})$ has a correspondence with a $\mathcal{N}=2$ SUSY extension model possessing self-dual or BPS structure.  Besides, the solutions of the Eqs. (\ref{1rd}) and (\ref{2rd}) correspond to the type $1/4$-BPS related to the nontrivial phase of the supersymmetric model  \cite{Queiruga}.}

{With the aim to study the properties of the BPS configurations we have used a rotationally symmetric {ansatz}. In such {ansatz} it is verified the total energy (\ref{en7}) of the self-dual configurations is proportional to the topological charge $N$ of the Skyrme field, thus, it is quantized. Then, we analyze the asymptotic behavior ($r\rightarrow\infty$) of the solutions by choosing a superpotential function that in such a limit behaves as $W(h)\approx h^{\sigma}/{\lambda^2}$. It has allowed to found two classes of self-dual profiles for the Skyrme field: the first class was obtained by considering $\sigma=2$ which provides solutions whose tail decays following an exponential-law $e^{-{\Lambda}r^2}$ with $\Lambda$ given in Eq. (\ref{expsq}). The second class occurs for $\sigma>2$, they are solutions whose tail decays following a power-law $r^{-\beta(\sigma)}$, with $\beta(\sigma)=2/(\sigma-2)${\color{black}, see Eq. (\ref{hsigS})}. For both classes of Skyrmions profiles, the respective  gauge fields possess an exponential-law type $e^{-\kappa r}$ (i.e., the Chern-Simons coupling constant becomes the gauge field mass) very similar to behavior found for such fields in Abelian Higgs models describing Abrikosov-Nielsen-Olesen vortices.}

{\color{black}Next, we dedicate our effort to solve numerically the differential equations describing the BPS configurations in order to attain the main properties or characteristics. For such a purpose we consider the superpotential defined by $W(h)=h^{\sigma}/{\lambda^2}$ and we study the solitons for $\sigma\geq 2$. It is shown the  soliton profiles  exhibit a compactonlike format for sufficiently large values of the electromagnetic coupling constant $g$. Further, the soliton solutions carry magnetic flux and possess nonzero total electric charge and, despite of both be proportional to the winding number $N$, they are nonquantized quantities  because the vacuum value  $a_\infty$ is a noninteger number [see Eqs. (\ref{flux}) and (\ref{cf2}), respectively]. However, it is shown numerically that for sufficiently large values of the electromagnetic coupling $g$  the vacuum value $a_\infty\rightarrow -1$, thus, both the quantities becomes effectively quantized, in accordance with previous investigations \cite{gladikowski,samoilenka,adam2,Casana01}.}

{\color{black}The more remarkable property is the emergence of the flipping of the magnetic field which implies in a localized magnetic flux inversion. This interesting feature emerges due to presence of both the  Maxwell and Chern-Simons terms in the BPS model (\ref{L03}). We emphasize such a feature is absent in the {\color{black}previous investigations of} gauged restricted baby Skyrme models with the Maxwell term  \cite{adam2} or the Chern-Simons term \cite{Casana01} solely .}

We now are investigating the existence of BPS solitons in a gauged baby Skyrme models into the presence of Lorentz violation. The results will be reported elsewhere.

\begin{acknowledgments}
{\color{black}This study was financed in part by the Coordena\c{c}\~ao de Aperfei\c{c}oamento de Pessoal de N\'{\i}vel Superior - Brasil (CAPES) - Finance Code 001. We thank also the Conselho Nacional de Desenvolvimento Cient{\'\i}fico e Tecnol\'ogico (CNPq), and the Funda\c{c}\~ao de Amparo \`a Pesquisa e ao Desenvolvimento Cient{\'\i}fico e Tecnol\'ogico do Maranh\~ao (FAPEMA) (Brazilian Government agencies). In particular, ACS, CFF and ALM thank the full support from CAPES. RC acknowledges the support from the grants
CNPq/306385/2015-5, CNPq/423862/2018-9 and FAPEMA/Universal-01131/17.}
\end{acknowledgments}

\end{document}